\documentclass{ieeetran}
\usepackage{graphicx}          
\usepackage{amssymb}
\usepackage{balance}
\usepackage{amsmath}
\usepackage{graphicx}
\usepackage{slashbox}
\usepackage{array}
\usepackage{color}
\usepackage{amsfonts}
\DeclareMathOperator{\sech}{sech}
\newtheorem{Theorem 1}{Theorem}
\newtheorem{Theorem 2}[Theorem 1]{Theorem}
\newtheorem{Theorem 3}[Theorem 1]{Theorem}
\newtheorem{Theorem 4}[Theorem 1]{Theorem}
\newtheorem{Theorem 5}[Theorem 1]{Theorem}
\newtheorem{Theorem 6}[Theorem 1]{Theorem}
\newtheorem{Theorem 7}[Theorem 1]{Theorem}
\newtheorem{Theorem 8}[Theorem 1]{Theorem}
\newtheorem{Assumption 1}{Assumption}
\newtheorem{Assumption 2}[Assumption 1]{Assumption}
\newtheorem{Assumption 3}[Assumption 1]{Assumption}
\newtheorem{Assumption 4}[Assumption 1]{Assumption}
\newtheorem{Assumption 5}[Assumption 1]{Assumption}
\newtheorem{Definition 1}{Definition}
\newtheorem{Remark 1}{Remark}
\newtheorem{Remark 2}[Remark 1]{Remark}
\newtheorem{Remark 3}[Remark 1]{Remark}
\newtheorem{Remark 4}[Remark 1]{Remark}
\newtheorem{Remark 5}[Remark 1]{Remark}
\newtheorem{Remark 6}[Remark 1]{Remark}
\newtheorem{Remark 7}[Remark 1]{Remark}
\newtheorem{Remark 8}[Remark 1]{Remark}
\newtheorem{Remark 9}[Remark 1]{Remark}
\newtheorem{Remark 10}[Remark 1]{Remark}
\newtheorem{Remark 11}[Remark 1]{Remark}
\newtheorem{Lemma 1}{Lemma}
\newtheorem{Lemma 2}[Lemma 1]{Lemma}
\newtheorem{Lemma 3}[Lemma 1]{Lemma}
\begin{document}
\title{Increasing sync rate of pulse-coupled oscillators via  phase response function design: theory and application to wireless networks}
\author{Yongqiang Wang, {\it Member, IEEE}, Felipe
N\'u\~nez, Francis J. Doyle III {\it Fellow, IEEE}\thanks{The work
was supported in part by U.S. Army Research Office through Grant
W911NF-07-1-0279, National Institutes of Health through Grant
GM078993, and the Institute for Collaborative Biotechnologies
through grant W911NF-09-0001 from the U.S. Army Research Office. The
content of the information does not necessarily reflect the position
or the policy of the Government, and no official endorsement should
be inferred.}
\thanks{Yongqiang Wang,  Francis J. Doyle III are with Department of Chemical Engineering, University of California, Santa
Barbara,  California USA. E-mail: wyqthu@gmail.com,
frank.doyle@icb.ucsb.edu. Felipe N\'u\~nez is with Department of
Electrical and Computer Engineering, University of California, Santa
Barbara,  California  USA. E-mail: fenunez@engineering.ucsb.edu} } \maketitle      

\begin{abstract}                          
This paper addresses the synchronization rate of weakly connected
pulse-coupled oscillators (PCOs). We prove that besides coupling
strength, the phase response function is also a determinant of
synchronization rate. Inspired by the result, we propose to increase
the synchronization rate of PCOs by designing the phase response
function. This has important significance in PCO-based clock
synchronization of wireless
 networks. By designing the phase response function, synchronization rate
is increased even under a fixed  transmission power. Given that
energy consumption in synchronization  is determined by the product
of synchronization time and transformation power, the new strategy
reduces energy consumption in clock synchronization. QualNet
 experiments  confirm the theoretical results.
\end{abstract}
\section{Introduction}
In recent years, synchronization of oscillating dynamical systems is
receiving increased attention. One particular class of oscillating
dynamical systems, pulse-coupled oscillators (PCOs), are of
considerable interest. `Pulse-coupled' means that oscillators
interact with each other using pulse-based communication, i.e.,
 they can achieve synchronization via the exchange of simple identical pulses. PCO has been used to
describe many biological synchronization phenomena such as the
flashing of fireflies, the contraction of cardiac cells, and the
firing of neurons \cite{mirollo:90}. Recently, with the progress of
ultra-wide bandwidth (UWB) impulse radio technology
\cite{Abdrabou:06}, the PCO based synchronization scheme has also
been applied to synchronize wireless networks \cite{hong:05,
Tyrrel:10, wernerallen:05, Pagliari:10,wang:12}. Since it is
implemented
 at the physical layer or MAC  (Media Access Control) layer, it eliminates the
high-layer intervention. Moreover, message exchanging in PCO-based
synchronization strategy is independent of the origin of the pulses,
which avoids requiring memory to store time information of other
nodes \cite{hong:05}. Therefore, the PCO-based synchronization
scheme has received increased attention in the communication
community recently.

Despite considerable work on synchronization conditions, there
remains a lack of research on the synchronization rate for PCOs,
especially for PCOs with a general coupling structure other than the
commonly studied all-to-all structure. The synchronization rate is
crucial in synchronization processes \cite{wang_automatica:11}. For
example, in the clock synchronization of wireless networks, the
synchronization rate is a determinant of the consumption in energy,
which is a precious system resource \cite{hong:05}.

This paper analyzes the synchronization rate of weakly connected
PCOs in the presence of combined global cues (also called leader, or
pinner in the language of pinning control \cite{Delellis:11}) and
local cues (alternatively, local coupling). The network structure is
considered because in the clock synchronization of wireless
networks, usually different time references are synchronized through
internal interplay between different nodes and external coordination
from a global time base such as GPS \cite{Kopetz:87}. Due to
pulsatile coupling, the synchronization rate of PCOs are very
difficult to study analytically \cite{mirollo:90}. Based on the
assumption of `weakly connected', we study the problem using phase
response functions. A phase response function describes the phase
correction of an oscillator induced by a pulse from
neighboring oscillators or external stimuli \cite{Izhikevich:07}. 
Under the assumption of weak  coupling, we transform the PCO model
into a simpler phase model, based on which we analyze the
synchronization rate of PCOs. In fact, we will prove that the
synchronization rate is determined not only by the strength of
global and local cues, but also by the phase response function. This
means that different phase response functions bring different
synchronization rates even when the coupling strength is fixed. This
has great significance for synchronization strategies such as the
clock synchronization of wireless  networks, where the phase
response function is a design parameter. By designing the phase
response function, we increase the synchronization rate under a
fixed coupling strength. Given that the total energy consumption in
synchronization is determined by the product of synchronization time
and transmission power (corresponding to coupling strength and
topology) \cite{hong:05,Barbarossa:07}, the new strategy reduces
energy consumption in  synchronization in that synchronization time
is reduced under a fixed transmission power. It is worth noting that
the assumption of weak coupling is well justified by biological
observations:  the amplitudes of postsynaptic potentials are around
0.1 mV, which is small compared with the amplitude of excitatory
postsynaptic potential necessary to discharge a quiescent cell
(around 20 mV) \cite{hoppensteadt:97}. In PCO-based wireless network
synchronization schemes, weak coupling is also necessary to
guarantee a robust synchronization \cite{hong:05}.

\section{Problem formulation and Model transformations}\label{sec:problem
formulation}

Consider a network of $N$ pulse-coupled oscillators, which will
henceforth be referred to as `nodes'. All oscillator nodes or a
portion of them can receive alignment/entrainment information from
an external global cue (also called leader, or pinner in the
language of pinning control \cite{Delellis:11}).

We denote the dynamics of the oscillator network as
\begin{equation}\label{eq:pulse_coupled_oscillators_original model}
\begin{aligned}
\dot{x}_g&=f_g(x_g)\\
\dot{x}_{i}&=f_i(x_i)+g_i\delta(t-t_g)+l\sum_{1\leq j\leq N,\,j\neq
i}a_{i,j}\delta(t-t_j)
\end{aligned}
\end{equation}
for $i=1,2,\ldots,N$, where $x_g\in[0,\,1]$ and $x_i\in [0,\,1] $
denote the states of the global cue and oscillator nodes,
respectively. $f_g$ and $f_i$ describe their dynamics. $g_i\geq0$
denotes the effect of the global cue's firing on oscillators $i$:
when  $x_g$ reaches 1 (at time instant $t_g$), it fires and returns
to 0, and at the same time increases oscillator $i$  by an amount
$g_i$. $l\geq0$ and $a_{ij}\in \{0,\,1\}$ denote the effect of
oscillator $j$'s firing on oscillator $i$: when  $x_j$ reaches 1 (at
time instant  $t_j$), it fires and resets to 0, and at the same time
pulls oscillator $i$ up by an amount $la_{i,j}$. The increased
amount is produced by dirac function $\delta(t)$, which is zero for
all $t$ except $t=0$ and satisfies
$\int_{-\infty}^{\infty}\delta(t)dt=1$.

\begin{Remark 1}
If $g_i$ (or $a_{i,j}$) is $0$, then oscillator $i$ is not affected
by the global cue (or oscillator $j$).
\end{Remark 1}
\begin{Assumption 1}
We assume $a_{i,j}=a_{j,i}$, which is common in wireless networks
\cite{Rappaport:02,Park:02}.
\end{Assumption 1}

\begin{Assumption 2}\label{assump:assumption 1}
We assume weak coupling \cite{hoppensteadt:97}, i.e., $g$ and $l$
satisfy $g\ll 1$ and $l\ll 1$.
\end{Assumption 2}

Assumption \ref{assump:assumption 1} follows from the fact that the
amplitudes of postsynaptic potentials measured in the soma of
neurons are far below
 the amplitude of the
mean excitatory postsynaptic potential necessary to discharge a
quiescent cell  \cite{hoppensteadt:97}, it is also required in
PCO-based wireless network synchronization strategies to ensure the
robustness of synchronization \cite{hong:05}.

Based on Assumption \ref{assump:assumption 1}, the system in
(\ref{eq:pulse_coupled_oscillators_original model}) can be described
by the following phase model using the classical  phase reduction
technique and phase averaging technique
\cite{Izhikevich:07,Vreeswijk:94}:

\begin{equation}\label{eq:pulse_coupled_oscillators_averaging}
\begin{aligned}
\dot{\theta}_g&=w_g\\
\dot{\theta}_{i}&=w_i+\frac{g_i}{T}Q_g(\theta_g-\theta_i)+\frac{l}{T}\sum_{1\leq
j\leq N,j\neq i}a_{i,j}Q_l(\theta_j-\theta_i)
\end{aligned}
\end{equation}
for $i=1,2,\ldots,N$, where $\theta_g \in [0,\,2\pi)$ and
$\theta_i\in[0,\,2\pi)$ denote the phases of the global cue and
oscillator  $i$, respectively. $Q_g(x)$ and $Q_l(x)$ are phase
response functions and are often referred to as phase response
curves in biological study. They are periodic functions with period
$2\pi$ \cite{Izhikevich:07,Vreeswijk:94}. $T$ is the period of the
global cue. $w_g$ and $w_i$ denote the natural frequencies of the
global cue and oscillator $i$, respectively.

\begin{Remark 1}
The transformation from (\ref{eq:pulse_coupled_oscillators_original
model}) to (\ref{eq:pulse_coupled_oscillators_averaging}) is a
standard practice in the study of weakly connected PCOs and it is
applicable to any limit-cycle oscillation function $f_i$ and $f_g$
\cite{Izhikevich:07}. The detailed procedure has been well
documented in \cite{Vreeswijk:94}, Chapter 9 of
\cite{hoppensteadt:97},
 and Chapter 10 of \cite{Izhikevich:07}.
\end{Remark 1}
\begin{Assumption 2}\label{assump:assumption 2}
In the paper, we assume that $Q_p\,(p=\{l,\,g\})$ satisfy the
following conditions:
\begin{equation}\label{eq:Assumptions on Q}
Q_p(0)=0;\:\forall x\in(-\pi,\pi),\frac{Q_p(x)}{x}>0;
Q_p(-x)=-Q_p(x)
\end{equation}
%
\end{Assumption 2}

\begin{Remark 2}
Assumption \ref{assump:assumption 2} gives an advance-delay phase
response function, which is common in biological oscillators
\cite{Izhikevich:07}. Moreover, given that in wireless networks, the
phase response function is a design parameter, Assumption
\ref{assump:assumption 2} will simplify analysis and design, and as
shown later, such phase response functions will also lead to good
synchronization properties.
\end{Remark 2}


Solving the first equation in
(\ref{eq:pulse_coupled_oscillators_averaging}) gives the dynamics of
the global cue $\theta_g=w_gt+\xi_g$, where the constant $\xi_g$
denotes the initial phase of the global cue. To study if  local
oscillators can be synchronized to the global cue, it is convenient
to study the phase deviation  of local oscillators from the global
cue. So we introduce the following change of variables:
\begin{equation}\label{eq:change to autonomous system}
\theta_i=\theta_g+\xi_i=w_gt+\xi_g+\xi_i
\end{equation}
Therefore $\xi_i\in[-\pi,\,\pi]$ denotes the phase deviation of the
$i$th oscillator from the global cue. Substituting (\ref{eq:change
to autonomous system}) into
(\ref{eq:pulse_coupled_oscillators_averaging}) yields the dynamics
of phase deviations $\xi_i$:
\begin{equation}\label{eq:pulse_coupled_oscillators_xi}
\dot{\xi}_{i}=\Delta_i-\frac{g_i}{T}Q_g(\xi_i)+\frac{l}{T}\sum_{1\leq
j\leq N,j\neq i}a_{i,j}Q_l(\xi_j-\xi_i)
\end{equation}
for $i=1,2,\ldots,N$,  where $\Delta_i=w_i-w_g$. In
(\ref{eq:pulse_coupled_oscillators_xi}), the oddness property of
function $Q_g$ is exploited.


\begin{Assumption 3}\label{assump:identical frequency}
In this paper, we assume $w_i=w_g$ is satisfied for all
$i=1,2,\ldots,N$, i.e., all the oscillators have the same natural
frequency as the global cue.
\end{Assumption 3}

Using Assumption \ref{assump:identical frequency},
(\ref{eq:pulse_coupled_oscillators_xi}) reduces to:
\begin{equation}\label{eq:dynamics under identical frequency}
\dot{\xi}_{i}=-\frac{g_i}{T}Q_g(\xi_i)+\frac{l}{T}\sum_{1\leq j\leq
N,j\neq i}a_{i,j}Q_l(\xi_j-\xi_i)
\end{equation}
for $i=1,2,\ldots,N$.

Thus far, by analyzing the properties of  (\ref{eq:dynamics under
identical frequency}), we can obtain the roles of global and local
cues as well as phase response functions in the synchronization of
PCO networks:
\begin{itemize}
\item Synchronization: If all $\xi_i$
asymptotically converge to $0$, then we have
$\theta_1=\theta_2=\ldots=\theta_N$ when time goes to infinity,
meaning that all the nodes are synchronized to the global cue.
\item Exponential bound on the synchronization rate: From dynamic systems theory \cite{Khalil:02}, the synchronization rate
 is determined by the rate at which $\xi_i$ decays
to $0$, namely, it
 can be measured by the maximal value of $\alpha$ $(\alpha>0)$
 satisfying the following inequality for some constant $C$:
\begin{equation}\label{eq:definition of sync rate}
\|\xi(t)\|\leq Ce^{-\alpha t}\|\xi(0)\|,\quad
\xi=\left[\begin{array}{cccc}\xi_1&\xi_2&\ldots&\xi_N\end{array}\right]^T
\end{equation}
where $\|\bullet\|$ is the Euclidean norm. A larger $\alpha$ leads
to a faster synchronization rate.
\end{itemize}

Assigning arbitrary orientation to each interaction, we can get the
$N\times M$ incidence matrix  $B$  ($M$ is the number of non-zero
$a_{i,j}\,(1\leq i\leq N,\,j<i)$, i.e., the number of interaction
edges) of the interaction \cite{Godsil:01}: $B_{i,j}=1$ if edge $j$
enters node $i$, $B_{i,j}=-1$ if edge $j$ leaves node $i$, and
$B_{i,j}=0$ otherwise. Then using graph theory, we can write
(\ref{eq:dynamics under identical frequency}) in a more compact
matrix form:
\begin{equation}\label{eq:matrix form}
\dot\xi=-\frac{1}{T}GQ_g(\xi)-\frac{l}{T}BQ_l\left(B^T\xi\right)
\end{equation}
where $\xi$ is given in (\ref{eq:definition of sync rate}),
$ G=\textrm{diag}\left\{g_1,\:g_2,\:\ldots,\:g_N\right\}$, and
$\textrm{diag}\{\bullet\}$ denotes a diagonal matrix with elements
$\{\bullet\}$ on the diagonal.


\section{Synchronization of Pulse Coupled Oscillators}\label{se:synchronization of pulse coupled
oscillators}

\subsection{When all  $\xi_i$ are within  $(-\frac{\pi}{2},\,\frac{\pi}{2})$}

\begin{Theorem 1}\label{theo:Theorem 1}
For the oscillator network in (\ref{eq:matrix form}), if all $\xi_i$
are within $[-\varepsilon,\,\varepsilon]$ for some
$\varepsilon\in[0,\,\frac{\pi}{2})$, then the network synchronizes
to the global cue when at least one $g_i$ is positive and the local
coupling topology $a_{i,j}$ is connected. Here `connected' means
that there is a multi-hop path (i.e., a sequence with nonzero values
$a_{i,m_1}$, $a_{m_1,m_2}$, $\ldots,$ $a_{m_{p-1},m_p}$,
$a_{m_p,j}$) from each node $i$ to every other node $j$.
\end{Theorem 1}
\begin{proof}
 We first prove that for any $0\leq \varepsilon<\frac{\pi}{2}$, when $\xi\in
[-\varepsilon,\,\varepsilon]\times
\ldots\times[-\varepsilon,\,\varepsilon]\triangleq[-\varepsilon,\,\varepsilon]^N$
where $\times$ is Cartesian product, they will remain in the
interval, i.e, $[-\varepsilon,\,\varepsilon]^N$ is positively
invariant for (\ref{eq:matrix form}). To this end, we only need to
check the direction of the vector field on the boundaries. If
$\xi_i=\varepsilon$, we have $0\leq\xi_i-\xi_j\leq 2\varepsilon<
\pi$ for $1\leq j\leq N$, so from (\ref{eq:dynamics under identical
frequency}) and the properties of phase response functions in
Assumption \ref{eq:Assumptions on Q}, $\dot{\xi}_i< 0$ holds. Hence
the vector field is pointing inward in the set, and no trajectories
can escape to values larger than $\varepsilon$. Similarly, we can
prove that when $\xi_i=-\varepsilon$, $\dot{\xi}_i> 0$ holds. Thus
the vector field is pointing inward in the set, and  no trajectories
can escape to values smaller than $-\varepsilon$. Therefore
$[-\varepsilon,\,\varepsilon]^N$ is positively invariant for any
$0\leq \varepsilon<\frac{\pi}{2}$.

Next we proceed to prove synchronization. Construct a Lyapunov
function as $ V=\frac{1}{2}\xi^T\xi $. $V$ is non-negative and will
be zero if and only if all $\xi_i$ are zero, meaning that all
oscillators are synchronized to the global cue.

Differentiating $V$ along the trajectories of (\ref{eq:matrix form})
yields
\begin{equation}\label{eq:lyapunov in theorem 1}
\begin{aligned}
\dot{V}=\xi^T\dot{\xi}&=-\frac{1}{T}\xi^TGS_1\xi-\frac{l}{T}\xi^TBS_2B^T\xi\\
&=-\frac{1}{T}\xi^T\left(GS_1+lBS_2B^T\right)\xi
\end{aligned}
\end{equation}
where $S_1$ and $S_2$ are given by
\begin{equation}\label{eq:S1}
S_1=\textrm{diag}\left\{\frac{Q_g(\xi_1)}{\xi_1},\,\frac{Q_g(\xi_2)}{\xi_2},\,\ldots,\,\frac{Q_g(\xi_N)}{\xi_N}\right\},
\end{equation}
\begin{equation}\label{eq:S2}
\begin{aligned}
&S_2= \textrm{diag}\left\{\frac{Q_l
((B^T\xi)_1)}{(B^T\xi)_1},\,\frac{Q_l((B^T\xi)_2)}{(B^T\xi)_2},\,\ldots,\,\frac{Q_l((B^T\xi)_M)}{(B^T\xi)_M}\right\}
\end{aligned}
\end{equation}
with $(B^T\xi)_i\,(1\leq i\leq M)$ denoting the $i$th element of the
$M\times 1$ dimensional vector $B^T\xi$.

According to dynamic systems theory \cite{Khalil:02}, if
$GS_1+lBS_2B^T$ in (\ref{eq:lyapunov in theorem 1}) is positive
definite, then  $\dot{V}$ is always negative when $\xi\neq0$ and $V$
will decay to zero exponentially, meaning that $\xi$ will converge
to zero and all oscillators are synchronized to the global cue.

Note that $(B^T\xi)_i\,(1\leq i\leq M)$ are in the form of
$\xi_m-\xi_n\,(1\leq m,n\leq N)$, it follows that $(B^T\xi)_i$ are
restricted to
 $[-2\varepsilon,\,2\varepsilon]$ when  all $\xi_j$ are in $[-\varepsilon,\,\varepsilon]$ for some $\varepsilon\in[0,\,\frac{\pi}{2})$.
 Given that in  $(-\pi,\,\pi)$, $Q_g(x)$ and $Q_l(x)$ satisfy $ \frac{Q_g(x)}{x}>
0,\: \frac{Q_l(x)}{x}> 0 $, it follows that $S_1$ and $S_2$ are
positive definite, and thus the following inequalities are satisfied
for some positive constants $\sigma_1$ and $\sigma_2$:
\begin{equation}\label{eq:sigma_1 and sigma_2}
\begin{aligned}
&S_1\geq \sigma_1 I, \quad S_2\geq \sigma_2 I,&\\
&\sigma_1=\min\limits_{-\varepsilon\leq
x\leq\varepsilon}\frac{Q_g(x)}{x},\quad
\sigma_2=\min\limits_{-2\varepsilon\leq
x\leq2\varepsilon}\frac{Q_l(x)}{x},\quad
\varepsilon\in[0,\,\frac{\pi}{2})&
\end{aligned}
\end{equation}

So we have $GS_1+lBS_2B^T\geq \sigma_1G+\sigma_2lBB^T$, which in
combination with (\ref{eq:lyapunov in theorem 1}) produces
\begin{equation}
\dot{V}\leq -\frac{1}{T}\xi^T\left(\sigma_1G+\sigma_2lBB^T
\right)\xi
\end{equation}
Next we prove that $\sigma_1G+\sigma_2lBB^T$ is positive definite,
which leads to $\dot V<0$ for $\xi\neq 0$.

It can be easily verified that $\sigma_1G+\sigma_2lBB^T$ is of the
following form:
\begin{equation}\label{eq:explicit formulation of Lambda}
\sigma_1G+\sigma_2lBB^T=\sigma_1{\rm
diag}\{g_1,\,g_2,\,\ldots,\,g_N\}+\sigma_2lL
\end{equation}
with $L\in\mathcal{R}^{N\times N}$ constructed as follows: for
$i\neq j$, its $(i,j)$th element is $-a_{i,j}$, for $i=j$, its
$(i,j)$th element is $\sum\limits_{ m=1, m\neq i}^N a_{i,m}$. Since
$\sigma_1$, $\sigma_2$, and $l$ are positive, and $g_i$, $a_{i,j}$
are non-negative, it follows from the Gershgorin Circle Theorem that
$\sigma_1G+\sigma_2lBB^T$ only has  non-negative eigenvalues
\cite{horn:85}. Next we prove its positive definiteness by excluding
$0$ as an eigenvalue.

Since  the topology of local coupling $a_{i,j}$ is connected,
$\sigma_1G+\sigma_2lBB^T$ is irreducible according to graph theory
\cite{horn:85}. This in combination with the assumption of at least
one non-zero $g_i$ guarantees that $\sigma_1G+\sigma_2lBB^T$ is
irreducibly diagonally dominant. So from Corollary 6.2.27 of
\cite{horn:85}, we know the determinant of $\sigma_1G+\sigma_2lBB^T$
is non-zero and hence 0 is not its eigenvalue. Therefore
$\sigma_1G+\sigma_2lBB^T$ is positive definite, and  $V$ will
converge to $0$.
\end{proof}
\subsection{When all  $\xi_i$ are within   $(-\pi,\,\pi)$ and the  maximal/minimal $\xi_i$ is outside
 $(-\frac{\pi}{2},\,\frac{\pi}{2})$}

\begin{Theorem 2}\label{theo:Theorem 2}
For the oscillator network in (\ref{eq:matrix form}), if all $\xi_i$
are within $[-\varepsilon,\varepsilon]$ for some
$\varepsilon\in[\frac{\pi}{2},\,\pi)$ and the maximal/minimal
$\xi_i$ is outside $(-\frac{\pi}{2},\,\frac{\pi}{2})$, then the
oscillator network will synchronize to the global cue when all nodes
are connected to the global cue, and the following relations are
satisfied:
\begin{equation}\label{eq:sync condition in theorem2}
\begin{aligned}
&g_{\min}> \frac{\sigma_4l\lambda_{\max}(BB^T)}{\sigma_3},&\\
&g_i\geq \frac{l}{\gamma_1}\sum_{1\leq j\leq N, j\neq
i}a_{i,j}\gamma_2,\quad i=1,2,\ldots,N&
\end{aligned}
\end{equation}
where $\lambda_{\max}$ denotes the maximal eigenvalue,
$g_{\min}=\min\{g_1,\,g_2,\,\ldots,\,g_N\}$, and
\begin{equation}\label{eq:gmin_sigma_3_sigma_4}
\begin{aligned}
& \sigma_3=\min\limits_{-\varepsilon\leq x\leq
 \varepsilon}\frac{Q_g(x)}{x},\,
\sigma_4=\max\limits_{-2\varepsilon\leq x \leq
2\varepsilon}\frac{-Q_l(x)}{x},&\\
&\gamma_1=\min\limits_{0\leq x\leq\varepsilon}Q_g(x),\,
\gamma_2=\max\limits_{0\leq
x\leq2\varepsilon}-Q_l(x),\,\varepsilon\in[\frac{\pi}{2},\,\pi)&
\end{aligned}
\end{equation}
\end{Theorem 2}
\begin{proof}
Following the line of reasoning of Theorem \ref{theo:Theorem 1}, we
can prove that if the second inequality in (\ref{eq:sync condition
in theorem2}) holds, then for any $\frac{\pi}{2}\leq
\varepsilon<\pi$, $[-\varepsilon,\,\varepsilon]^N$ is positively
invariant for (\ref{eq:matrix form}), i.e., for
$\xi\in[-\varepsilon,\,\varepsilon]^N$, it will always remain in the
interval. Next we proceed to prove synchronization.

Choose the same Lyapunov function $V$ as the proof of Theorem
\ref{theo:Theorem 1}. Then  we have
\begin{equation}\label{eq:lyapunov in theorem 2}
\begin{aligned}
\dot{V}=\xi^T\dot{\xi}&=-\frac{1}{T}\xi^TGS_1\xi-\frac{l}{T}\xi^TBS_2B^T\xi\\
&=-\frac{1}{T}\xi^T\left(GS_1+lBS_2B^T\right)\xi
\end{aligned}
\end{equation}
where $S_1$ and $S_2$ are given in (\ref{eq:S1}) and (\ref{eq:S2}),
respectively.

When the maximal/minimal $\xi_i$ is outside
$(-\frac{\pi}{2},\,\frac{\pi}{2})$,  $\xi_m-\xi_n\,(1\leq m,n\leq
N)$ may be outside   $(-\pi,\, \pi)$. So in (\ref{eq:S1}) and
(\ref{eq:S2}), the domain of $Q_g(x)$ is within $(-\pi,\,\pi)$, on
which $Q_g(x)$ satisfies $Q_g(x)/x> 0$, and the domain of $Q_l(x)$
is not restricted to  $(-\pi,\,\pi)$, outside of which, $Q_l(x)/x$
may be positive or negative. Therefore $S_1$ is still positive
definite, but
$S_2$ may be positive definite, negative definite or indefinite. 
From the definition of $\sigma_3$ and $\sigma_4$ in
(\ref{eq:gmin_sigma_3_sigma_4}), we have:
\[
GS_1\geq \sigma_3g_{\min},\quad -lBS_2B^T\leq
\sigma_4l\lambda_{\max}(BB^T)
\]
Notice that  $Q_l(x)$ is periodic with period $2\pi$, and
$\frac{Q_l(x)}{x}> 0$  holds for all $-\pi < x < \pi$, we know for
any $\frac{\pi}{2}\leq \varepsilon<\pi$, if
$x_0\in[-2\varepsilon,\,-\pi]$, then $Q_l(x_0)=Q_l(2\pi+x_0)\geq 0$
holds since $2\pi+x_0$ resides in the interval
$[2(\pi-\varepsilon),\,\pi]$.
 Thus it follows  $\frac{Q_l(x_0)}{x_0}=\frac{Q_l(x_0+2\pi)}{x_0}\leq0$, which means $\sigma_4\geq0$.

Therefore, (\ref{eq:sync condition in theorem2}) guarantees the
positive definiteness of $GS_1+lBS_2B^T$, and hence the
synchronization of the oscillators to the global cue.
\end{proof}

\begin{Remark 5}
Theorem \ref{theo:Theorem 2} indicates that when the maximal/minimal
phase difference is outside $(-\frac{\pi}{2},\,\frac{\pi}{2})$, all
oscillators have to connect to the global cue  to ensure
synchronization to the global cue. This is consistent with existing
results which have shown that for some initial conditions (even with
measure zero), PCOs cannot be synchronized by local coupling
\cite{mirollo:90}. In fact, most of the existing results on PCOs are
based on all-to-all connection, which amounts to $g_{\min}>0$.
\end{Remark 5}

\begin{Remark 6}
Theorem \ref{theo:Theorem 2} reveals that a strong local cue does
not necessarily benefit synchronization when phase difference is
outside $(-\frac{\pi}{2},\,\frac{\pi}{2})$. This is also consistent
with \cite{Monzon:05} which  shows that synchronization may not be
achieved despite arbitrarily strong local coupling.
\end{Remark 6}

\section{Synchronization Rate of Pulse Coupled Oscillators}\label{se:synchronization of pulse coupled
oscillators}

Based on a similar derivation, we can get a bound on the exponential
synchronization rate:
\begin{Theorem 3}\label{theo:Theorem 3}
For the oscillator network in (\ref{eq:matrix form}), define
$\sigma_1$,
 $\sigma_2$ as in (\ref{eq:sigma_1 and sigma_2}), and $\sigma_3$,
 $\sigma_4$ as in (\ref{eq:gmin_sigma_3_sigma_4}), then
\begin{itemize}
\item when all $\xi_i$ are within $[-\varepsilon,\varepsilon]$ for
some  $\varepsilon\in[0,\,\frac{\pi}{2})$ and the conditions in
Theorem 1 hold, the synchronization rate is no worse than
\begin{equation}\label{eq:alpha_1}
\begin{aligned}
\alpha_1&=\min_{\xi}\left\{ \xi^T
\left(\sigma_1G+\sigma_2lBB^T\right)\xi
/(\xi^T\xi)\right\}/T\\
&=\lambda_{\min}\left(\sigma_1G+\sigma_2lBB^T\right)/T
\end{aligned}
\end{equation}
\item when the maximal/minimal $\xi_i$ is outside $(-\frac{\pi}{2},\,\frac{\pi}{2})$ and the conditions in Theorem 2 hold, the synchronization rate is no worse than
\begin{equation}\label{eq:alpha_2}
\alpha_2=\left(\sigma_3
g_{\min}-\sigma_4l\lambda_{\max}(BB^T)\right)/T
\end{equation}
\end{itemize}
\end{Theorem 3}

\begin{proof}
First consider the case that  all $\xi_i$ are within
$[-\varepsilon,\varepsilon]$ for some
$\varepsilon\in[0,\,\frac{\pi}{2})$. From (\ref{eq:lyapunov in
theorem 1}), it follows
\begin{equation}
\dot{V}\leq -2\alpha_1\frac{\xi^T\xi}{2}=-2\alpha_1 V
\end{equation}
with $\alpha_1$ defined in (\ref{eq:alpha_1}), which further means
that
\begin{eqnarray}
V(t)\leq C^2e^{-2\alpha_1t}V(0)\Rightarrow \|\xi(t)\|\leq C
e^{-\alpha_1t}\|\xi(0)\|
\end{eqnarray}
holds for some positive constant $C$. Thus the synchronization rate
is no worse than $\alpha_1$ in (\ref{eq:alpha_1}).

Similarly, we can prove from (\ref{eq:lyapunov in theorem 2}) that
when the maximal/minimal $\xi_i$  is outside
$(-\frac{\pi}{2},\,\frac{\pi}{2})$, the following relationship
\begin{eqnarray}
V(t)\leq C^2e^{-2\alpha_2t}V(0)\Rightarrow \|\xi(t)\|\leq C
e^{-\alpha_2t}\|\xi(0)\|
\end{eqnarray}
holds for some positive constant $C$. Thus the synchronization rate
is no worse than $\alpha_2$ in (\ref{eq:alpha_2}).
\end{proof}
\begin{Remark 4}
When $Q_g$ and $Q_l$ are sinusoidal functions, all $\xi_i$ are
constrained in the interval $(-\frac{\pi}{2},\frac{\pi}{2})$, and
there is no global cue ($G=0$), using $\bar\theta=\sum\limits_{1\leq
i\leq N}\frac{\theta_i}{N}$ as reference, we can define $\xi_i$ as
$\xi_i=\theta_i-\bar\theta$. Since $\xi^T{\bf 1}=0$ holds for ${\bf
1}=[1,1,\ldots,1]^T$, the constraint $\xi^T{\bf 1}=0$ is added to
the optimization $\min\limits_\xi\left\{ \xi^T
\left(\sigma_1G+\sigma_2lBB^T\right) \xi /(\xi^T\xi)\right\}/T$ in
(\ref{eq:alpha_1}). Given that $G=0$ and $BB^T$ is the Laplacian
matrix of interaction graph and hence has eigenvector ${\bf 1}$ with
associated eigenvalue $0$ \cite{horn:85}, $\lambda_{\min}$ in
(\ref{eq:alpha_1}) reduces to the second smallest eigenvalue, which
is the same as the convergence rate in section IV of \cite{Chung:10}
obtained using contraction analysis.
\end{Remark 4}

From application point of view, it is important to analyze how
synchronization rate is affected by the phase response function and
the strengths of global and local cues. According to
(\ref{eq:gmin_sigma_3_sigma_4}) and (\ref{eq:alpha_2}), it is clear
that the synchronization rate increases with an increase in
 $g_{\min}$ and $\frac{Q_g(x)}{x}$. But how the phase response function and the strength
of the global cue affect the  synchronization rate when all $\xi_i$
are within $(-\frac{\pi}{2},\,\frac{\pi}{2})$ is not clear. (In this
case, $g_{\min}$ may be zero since some oscillators may not be
connected to the global cue.) In fact, we can prove that in this
case the synchronization rate also increases with an increase in
$\frac{Q_g(x)}{x}$ and the strength of the global cue:
\begin{Theorem 4}\label{theo:Theorem 4}
The synchronization rate of (\ref{eq:matrix form}) increases with an
increase in the strength  of the global cue. It also increases with
an increase in $\frac{Q_g(x)}{x}$.
\end{Theorem 4}
\begin{proof}
As analyzed in the paragraph above Theorem \ref{theo:Theorem 4}, we
only need to prove the statement
 when all $\xi_i\in[-\varepsilon,\varepsilon]$ for
some  $\varepsilon\in[0,\,\frac{\pi}{2})$, i.e., $\alpha_1$  in
(\ref{eq:alpha_1}) is an increasing function of $g_i$ and
$\frac{Q_g(x)}{x}$. Recall from (\ref{eq:explicit formulation of
Lambda}) that $\sigma_1G+\sigma_2lBB^T$ is an irreducible matrix
with non-positive off-diagonal elements, so there exists a positive
$\mu$ such that $\mu I-(\sigma_1G+\sigma_2lBB^T)$ is an irreducible
non-negative matrix. Therefore, $\lambda_{\max}\left(\mu
I-(\sigma_1G+\sigma_2lBB^T)\right)$ is the  Perron-Frobenius
eigenvalue of $\mu I-(\sigma_1G+\sigma_2lBB^T)$ and is positive
\cite{horn:85}. Given that for any $1\leq i\leq N$,
$\mu-\lambda_i(\sigma_1G+\sigma_2lBB^T)$ is an eigenvalue of matrix
$\mu I -(\sigma_1G+\sigma_2lBB^T)$ where $\lambda_i$ denotes the
$i$th eigenvalue, we have
\[
\mu-\lambda_{\min}(\sigma_1G+\sigma_2lBB^T)=\lambda_{\max}\left(\mu
I -(\sigma_1G+\sigma_2lBB^T)\right)
\]
i.e.,
\[
\begin{aligned}
\alpha_1&=\frac{\lambda_{\min}(\sigma_1G+\sigma_2lBB^T)}{T}&\\
&=\frac{\mu-\lambda_{\max}\left(\mu I
-(\sigma_1G+\sigma_2lBB^T)\right)}{T}&
\end{aligned}
\]

Given that the largest eigenvalue (also called the Perron-Frobenius
eigenvalue) of $\mu I-(\sigma_1G+\sigma_2lBB^T)$ is an increasing
function of any of its diagonal
 element \cite{horn:85}, which is a decreasing function of $g_i$ and
 $\frac{Q_g(x)}{x}$, it follows  that $\lambda_{\max}\left(\mu
I -(\sigma_1G+\sigma_2lBB^T)\right)$ is a decreasing function of
both $g_i$ and $\frac{Q_g(x)}{x}$, meaning that $\alpha_1$ is an
increasing function of  $g_i$ and  $\frac{Q_g(x)}{x}$.
\end{proof}
\begin{Remark 3}\label{remark:affects of local cue}
The role of the local cue is not discussed in Theorem
\ref{theo:Theorem 4}. In fact, the role of the local cue depends on
the value of $\xi_i$:  when all $\xi_i$ are within
$(-\frac{\pi}{2},\,\frac{\pi}{2})$, $S_2$ in (\ref{eq:S2}) is
positive definite,
 so $\xi^TBS_2B^T\xi$ in (\ref{eq:lyapunov in
theorem 1}) is positive, meaning that the local cue will increase
the synchronization rate. Whereas when the maximal/minimal $\xi_i$
is outside of the interval $(-\frac{\pi}{2},\,\frac{\pi}{2})$, $S_2$
in (\ref{eq:S2}) can be positive semi-definite, negative
semi-definite or indefinite, $\xi^TBS_2B^T\xi$ in (\ref{eq:lyapunov
in theorem 2}) can be positive, negative or zero, thus the local cue
may increase, decrease or have no influence on the synchronization
rate. This conclusion is confirmed by QualNet experiments in Sec.
\ref{sec:QuanNet}.
\end{Remark 3}
\begin{Remark 4}\label{remark: remark 4}
From Theorem \ref{theo:Theorem 3} and Theorem \ref{theo:Theorem 4},
one can see that in addition to the strength of coupling, i.e.,
$g_i$ and  $l$, the phase response function  $Q_g$ also influences
the synchronization rate. This  has significant ramifications for
 the clock synchronization of wireless networks
using PCO-based strategies \cite{hong:05,Pagliari:10,Barbarossa:07},
where the phase response function is a design parameter: the
synchronization rate can be increased by choosing appropriate phase
response functions, even with transmission power (corresponding to
coupling strength and topology) fixed, therefore leading to a
reduced energy consumption. This will be addressed in Sec.
\ref{se:design of the PRC}.
\end{Remark 4}

\section{Design of Phase Response Functions}\label{se:design of the PRC}

As stated in Sec. \ref{se:synchronization of pulse coupled
oscillators}, the phase response function is an important
determinant of the synchronization rate of PCOs. This has important
ramifications for PCO-based synchronization strategies of wireless
networks, where the phase response function is a design parameter.

PCO-based synchronization strategies are attracting increased
 attention in the communications literature \cite{hong:05,Pagliari:10,wang:12,Barbarossa:07}.  As with most synchronization strategies, a network using a PCO-based
synchronization strategy makes a distinction between an acquisition
stage where the network synchronizes and the communication stage
where nodes transmit and receive data \cite{Rappaport:02}. In
PCO-based synchronization strategies, every node of the network acts
as a PCO, nodes interact through transmitting replicas of a pulse
signal, which can be a monocycle pulse in a UWB network
\cite{hong:05} or preambles in IEEE 802.11 networks
\cite{Tyrrell:06}. PCO-based strategies have many advantages over
conventional synchronization strategies \cite{hong:05}: they are
implemented at the physical layer or MAC layer, which eliminates the
high-layer intervention; their message exchanging is independent of
the origin of the signals, which avoids requiring memory to store
time information of other nodes.

In all existing PCO-based synchronization strategies, the oscillator
model is directly adopted from a biological source, leading to a
fixed phase response function. Our finding suggests that even under
a fixed coupling strength, one can increase the synchronization rate
by designing the phase response function. This can reduce energy
consumption in clock synchronization since the total energy
consumption in a synchronization process is determined by the
product of transmission power (corresponding to coupling strength
and topology) and the time to synchronization. Next, we show that by
designing the phase response function, we can indeed increase the
synchronization rate.

We focus on a class of phase response functions in $\tanh$ form, for
reasons outlined below:
\begin{equation}\label{eq:phase response_F}
Q(x)=\frac{\tanh(x/\varepsilon)}{\tanh(\pi/\varepsilon)}-\frac{x}{\pi},\quad\textrm{when}\quad
x\in[-\pi,\,\pi]
\end{equation}
where $\varepsilon>0$ is a free parameter and will be designed to
achieve a faster synchronization rate.

\begin{figure}[!hbp]
\begin{center}
  \includegraphics[width=0.8\columnwidth]{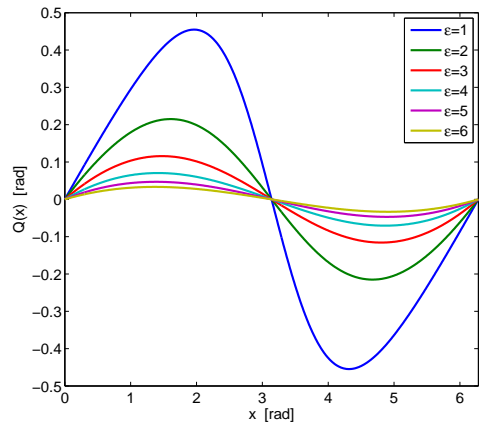}
 \caption{The shape of $\tanh$ type advance-delay
phase response functions in equation (\ref{eq:phase response_F}).}
    \label{fg:PRC}
\end{center}

\end{figure}

Fig. \ref{fg:PRC} gives the plot of $Q(x)$ in (\ref{eq:phase
response_F}). Since $Q(x)$ is $2\pi$-periodic, only its value in the
interval $[0,\,2\pi]$ is plotted. $Q(x)$ is an advance-delay phase
response function, i.e., external pulsatile input either delays or
advances an oscillator's phase, depending upon the timing of input.
Advance-delay phase response functions have been widely used in the
 biology community: they can well
 characterize the dependence of neurons' response to small depolarizations, i.e., an excitatory
 postsynaptic potential (EPSP) received after the refractory period delays the firing of the next spike,
 while an EPSP received at a later time advances the firing. The most widely-used neuron
 model, i.e., the
Hodgkin-Huxley model, also has advance-delay phase response
functions \cite{Ermentrout:96}.
From Theorem \ref{theo:Theorem 4}, we know in addition to the
strength of the global and local cues,  the phase response function
of the global cue also determines the synchronization rate: the
larger $ \frac{Q_g(x)}{x}$ is, the faster the synchronization rate.
In the following, we will show that the synchronization rate can be
increased by designing $\varepsilon$, a parameter in the phase
response function.
%
\begin{Theorem 5}\label{theo:Theorem 5}
For the pulse-coupled oscillator network  in (\ref{eq:matrix form})
and the phase response function of the global cue in the form of
(\ref{eq:phase response_F}), if all $\xi_i$ are within $(-\pi,\pi)$,
the synchronization rate increases with a decrease in $\varepsilon$,
no matter whether the maximal/minimal $\xi_i$ is within or outside
$(-\frac{\pi}{2},\frac{\pi}{2})$.
\end{Theorem 5}
\begin{proof}
According to Theorem \ref{theo:Theorem 4}, the synchronization rate
increases with an increase in $\frac{Q_g(x)}{x}$. So  we only need
to prove that $\frac{Q_g(x)}{x}$ increases with a decrease in
$\varepsilon$ for $x\in(-\pi,\,\pi)$.

Using the $\tanh$ type phase response function, we have
\begin{equation}\label{eq:Q over x in theorem 5}
\frac{Q_g(x)}{x}=\frac{\tanh(x/\varepsilon)}{x\tanh(\pi/\varepsilon)}-\frac{1}{\pi},\quad
x\in(-\pi,\,\pi)
\end{equation}
Since $\frac{Q_g(x)}{x}$ in (\ref{eq:Q over x in theorem 5}) is a
smooth function of $\varepsilon$, we can calculate its derivative
with respect to $\varepsilon$:
\begin{equation}
\frac{d(\frac{Q_g(x)}{x})}{d\varepsilon}=\frac{\pi\sech^2(\frac{\pi}{\varepsilon})\tanh(\frac{x}{\varepsilon})-x\sech^2(\frac{x}{\varepsilon})\tanh(\frac{\pi}{\varepsilon})}{x\varepsilon^2\tanh^2(\frac{\pi}{\varepsilon})}
\end{equation}
To prove that $\frac{Q_g(x)}{x}$ increases with a decrease in
$\varepsilon$, we need to prove that
$\frac{d(\frac{Q_g(x)}{x})}{d\varepsilon}$ is negative. Since
$\varepsilon^2\tanh^2(\frac{\pi}{\varepsilon})$ is positive, we only
need to prove that (\ref{eq:numerator}) is negative for $-\pi < x <
\pi$:
\begin{equation}\label{eq:numerator}
\frac{1}{x}\left(\pi\sech^2(\frac{\pi}{\varepsilon})\tanh(\frac{x}{\varepsilon})-x\sech^2(\frac{x}{\varepsilon})\tanh(\frac{\pi}{\varepsilon})\right)
\end{equation}
Using properties of hyperbolic functions, we can rewrite
(\ref{eq:numerator}) as follows:
\begin{equation}
\begin{aligned}
&\frac{4}{\left(e^{\frac{\pi}{\varepsilon}}+e^{\frac{-\pi}{\varepsilon}}\right)^2\left(e^{\frac{x}{\varepsilon}}+e^{\frac{-x}{\varepsilon}}\right)^2}&\\
&\times\
\frac{\pi\left(e^{\frac{2x}{\varepsilon}}-e^{\frac{-2x}{\varepsilon}}\right)-x\left(e^{\frac{2\pi}{\varepsilon}}-e^{\frac{-2\pi}{\varepsilon}}\right)}{x}&
\end{aligned}
\end{equation}
hence the problem is reduced to proving the negativity of
$f(x,\varepsilon)$ in (\ref{eq:exponentional expression in Theorem
5}) for $-\pi < x< \pi$.
\begin{equation}\label{eq:exponentional expression in Theorem 5}
f(x,\varepsilon)\triangleq\frac{\pi\left(e^{\frac{2x}{\varepsilon}}-e^{\frac{-2x}{\varepsilon}}\right)-x\left(e^{\frac{2\pi}{\varepsilon}}-e^{\frac{-2\pi}{\varepsilon}}\right)}{x}
\end{equation}

Using a Taylor expansion, equation (\ref{eq:exponentional expression
in Theorem 5}) can be further rewritten as

\begin{equation}
\begin{aligned}
f(x,\varepsilon)&=\frac{1}{x}\left[
2\pi\left(\frac{2x}{\varepsilon}+\frac{(\frac{2x}{\varepsilon})^3}{3!}+\frac{(\frac{2x}{\varepsilon})^5}{5!}+\ldots
\right)\right.\\
&\left.\qquad\quad
-2x\left(\frac{2\pi}{\varepsilon}+\frac{(\frac{2\pi}{\varepsilon})^3}{3!}+\frac{(\frac{2\pi}{\varepsilon})^5}{5!}+\ldots
\right)\right]\\
&=\frac{2}{x}\left(\frac{\pi(\frac{2x}{\varepsilon})^3-x(\frac{2\pi}{\varepsilon})^3}{3!}+\frac{\pi(\frac{2x}{\varepsilon})^5-x(\frac{2\pi}{\varepsilon})^5}{5!}
+\ldots \right)
\end{aligned}
\end{equation}
which is negative for all $x$ in $-\pi<x<\pi$.

So $\frac{(\frac{Q_g(x)}{dx})}{d\varepsilon}$ is negative, thus
$\frac{Q_g(x)}{x}$ increases with a decrease in $\varepsilon$, which
completes the proof.
\end{proof}

\section{QualNet Experiments}\label{sec:QuanNet}

We use a high-fidelity network evaluation tool (QualNet) to
illustrate the proposed strategy. QualNet is a commercial  network
platform that has been widely used to predict the performance of
MANETs, satellite networks and sensor networks, among others
\cite{QualNet:08}.

\begin{figure}[!hbp]
\begin{center}
  \includegraphics[angle=270, width=\columnwidth]{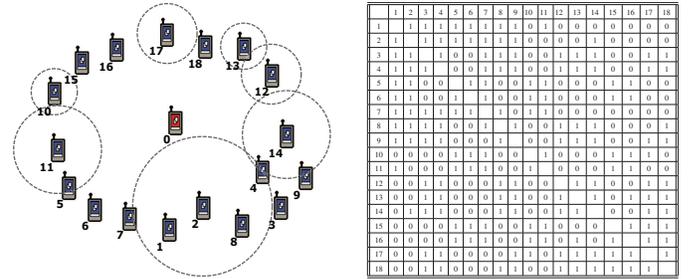}
    \caption{Schematic of the network (left) and interaction matrix (right) used in the QualNet implementation.
    In the interaction matrix, `1' denotes interaction between two nodes, and `0' denotes no interaction.}
    \label{fg:QualNet}
\end{center}
\end{figure}

In the implementation, we constructed a wireless network composed of
19 nodes (including 1 global cue). Each node has a counter as
 clock and
stores the phase response function (as shown in Fig. \ref{fg:PRC})
in a lookup table. Upon receiving a pulse, a node shifts its phase
by an amount determined by its current time and the phase response
function in the lookup table. The structure of the network is
illustrated in Fig. \ref{fg:QualNet}, where node number 0 is the
global cue. A broadcasting-based MAC layer protocol is adopted to
establish the pulse based communication between different nodes,
which is represented by the circles in Fig. \ref{fg:QualNet}.
Although a broadcasting scheme is used, the communication in the
network is not all-to-all due to limited transmission range. The
interaction topology is also illustrated in Fig. \ref{fg:QualNet},
which can be verified to be a connected graph.

We first considered the case where  $\xi_i\in(-\frac{\pi}{2},
\frac{\pi}{2})$. From Theorem \ref{theo:Theorem 1}, all nodes can
synchronize to the global cue if at least one $g_i$ is non-zero. To
confirm the prediction, we only connected oscillator $1$
 to the global cue with $g_1=0.01$. We set the natural frequency to $2\pi$, the strength of local coupling to $l=0.01$, and
implemented the PCO network under different phase response functions
in (\ref{eq:phase response_F}), i.e., different pairs of
$\varepsilon_g$ ($\varepsilon$ in $Q_g$) and $\varepsilon_l$
($\varepsilon$ in $Q_l$). The network was synchronized, confirming
Theorem \ref{theo:Theorem 1}. To illustrate our
phase-response-function based design strategy, we recorded the
average time to synchronization from 100 runs for each pair of
$\varepsilon_g$ and $\varepsilon_l$. In each run, we set the initial
value of the global cue as $0$ and chose the initial values of local
nodes randomly from a uniform distribution on
$(-\frac{\pi}{2},\,\frac{\pi}{2})$.  The times to synchronization
are given by the first element of each 2-tuple in Table I. We can
see that the synchronization rate increases with a decrease in
$\varepsilon_g$, which confirms the theoretical results in Sec.
\ref{se:design of the PRC}.
The total energy consumption in the synchronization process is also
recorded and averaged over the 100 runs. The results are given
 by the second element of each 2-tuple in Table I. With a decrease
in $\varepsilon_g$, the energy consumption indeed decreases, which
confirms the effectiveness of our design methodology in Sec.
\ref{se:design of the PRC}.
\begin{table*}[htb]\label{ta:tableQualNet_time1}
\begin{minipage}[t]{\linewidth}
\centering
 \caption{Time to synchronization [s]($1$st element of each 2-tuple) and energy consumption [$10^{-3}$J]($2$nd element of each 2-tuple)
 under different phase response functions ($\xi_i\in(-\frac{\pi}{2},\,\frac{\pi}{2})$, $l=g_1=0.01$, $g_2=g_3=\ldots=g_N=0$)}
\begin{tabular}{||c|c|c|c|c|c|c||}
\hline\hline
$\varepsilon_g$ $\backslash$ $\varepsilon_l$& $0.05$ & $0.1$ & $0.2$
&
 $0.4$  &  $0.8$& $1.6$
\\
\hline $0.4$&  (37.23, 729.39)  & (36.76, 719.99) & (36.55, 715.79)
& (36.23, 709.05) & (35.83, 701.39) & (43.37, 852.19)
\\
\hline $0.8$&  (37.88, 742.39)  & (37.40, 732.79)  & (37.88, 742.39)
& (36.73, 719.39)  & (36.44, 713.59)& (44.04, 865.59)
\\
\hline $1.6$&   (38.17, 748.19) & (37.90, 742.79) &  (38.38, 752.39)
& (37.62, 737.19) & (36.83, 721.43)& (45.76, 899.99)
\\
\hline\hline
\end{tabular}
\end{minipage}
\end{table*}

Using the same setup, we also implemented the network when the
maximal/minimal $\xi_i$ is outside $(-\frac{\pi}{2}, \,
\frac{\pi}{2})$. We ran the implementation for 100 times and each
time chose the initial values of $\xi_i$ randomly  from a uniform
distribution on $(-\pi,\,\pi)$. $28$ of the 100 runs were
unsynchronized, confirming Theorem \ref{theo:Theorem 2} that all
oscillators have to be connected to the global cue to guarantee
synchronization. So we made $g_1=g_2=\ldots=g_N=0.01$ and re-ran the
implementation under different phase response functions. The results
are given in Table II. With a decrease in $\varepsilon_g$, the
energy consumption indeed decreases, confirming the effectiveness of
our design methodology in Sec. \ref{se:design of the PRC}. Moreover,
using the same coupling strength, we also implemented the network
under Peskin's phase response function used in \cite{hong:05}, and
obtained a (synchronization time, energy consumption) 2-tuple as
$(25.24,\,489.59)$. Since it is larger than the smallest energy
consumption in Table II, which is obtained under the same coupling
strength, this confirms that by tuning the parameter $\varepsilon_g$
in phase response function, energy consumption can indeed be
reduced.
\begin{table*}[htb]\label{ta:tableQualNet_time1}
\begin{minipage}[t]{\linewidth}
\centering
 \caption{Time to synchronization [s]($1$st element of each 2-tuple) and energy consumption [$10^{-3}$J]($2$nd element of each 2-tuple)
 under different phase response functions ($\xi_i\in(-\pi,\,\pi)$, $l=g_1=g_2=\ldots=g_N=0.01$)}
\begin{tabular}{||c|c|c|c|c|c|c||}
\hline\hline
$\varepsilon_g$ $\backslash$ $\varepsilon_l$ & $0.05$ &
$0.1$ & $0.2$ &
 $0.4$  &  $0.8$& $1.6$
\\
\hline $0.4$&  (22.93, 443.39)  & (23.17, 448.19) & (23.14, 447.59)
& (22.58, 436.39) & (21.53, 415.39) & (22.44, 433.59)
\\
\hline $0.8$&  (24.95, 483.79)  & (25.21, 488.99)  & (25.36, 491.99)
& (24.23, 469.39)  & (23.63, 457.39)& (24.34, 471.59)
\\
\hline $1.6$&   (30.14, 587.59) & (31.92, 623.19) &  (31.75, 619.79)
& (30.35, 519.79) & (28.15, 547.79)& (29.09, 566.59)
\\
\hline\hline
\end{tabular}
\end{minipage}
\end{table*}

Setting $g_1=g_2=\ldots=g_N=g$, $\varepsilon_g=0.4$, and
$\varepsilon_l=0.05$, we also implemented the network under
different strengths of global and local cues, i.e., different pairs
of $g$ and $l$. For each pair of $g$ and $l$, we ran the
implementation for 100 times, and each time we chose the initial
values of $\xi_i$ randomly from a uniform distribution on
$(-\pi,\,\pi)$. The average time to synchronization is given by the
first element of each 2-tuple in Table III. From Table III, we can
see that a larger $g$ indeed leads to a faster synchronization rate
(a smaller synchronization time), whereas a larger $l$ does not
necessarily bring a faster synchronization rate. A larger $l$ may
even desynchronize the network when $g$ is small (as illustrated by
the last two elements of the first row). This confirms the
analytical results in Remark \ref{remark:affects of local cue},
which state that the local cue may increase or decrease the
synchronization rate. The same conclusion can be drawn for energy
consumption, which is given by the second element of each 2-tuple in
Table III.
\begin{table*}[htb]\label{ta:tableQualNet_time1}
\begin{minipage}[t]{\linewidth}
\centering
 \caption{Time to synchronization [s] ($1$st element of each 2-tuple) and energy consumption [$10^{-3}$J] ($2$nd element of each 2-tuple)
 under different strengths of global and local cues ($\xi_i\in(-\pi,\,\pi)$, $g_1=g_2=\ldots=g_N=g$)}
\begin{tabular}{||c|c|c|c|c|c|c||}
\hline\hline
$g$ $\backslash$ $l$& $0.01$ & $0.02$ & $0.03$ &
 $0.04$  &  $0.05$& $0.06$
\\
\hline $0.01$    &  (22.93, 443.39)  & (23.21, 448.99) & (26.37,
512.19) & (27.26, 529.99) & (no sync, $-$) & (no sync, $-$)
\\
\hline $0.02$    &  (17.49, 334.59)  & (18.90, 362.79)  & (22.03,
425.39) & (24.60, 476.79)  & (24.38, 472.39)& (21.39, 412.59)
\\
\hline $0.03$    &   (14.18, 268.39) & (14.99, 284.59) &  (18.03,
345.39) & (19.93, 383.39) & (20.35, 391.79)& (19.91, 382.99)
\\
\hline\hline
\end{tabular}
\end{minipage}
\end{table*}
\section{Conclusions}\label{se:conclusions}
The synchronization rate of pulse-coupled oscillators is analyzed.
It is proven that in addition to the strengths of global and local
cues, the phase response function also determines the
synchronization rate. This inspires us to increase the
synchronization rate by choosing an appropriate phase response
function when the phase response function is a design parameter. An
application is the clock synchronization of wireless networks, to
which pulse-coupled synchronization strategies have been
successfully applied. By exploiting the freedom in the phase
response function, we give a new design methodology for
pulse-coupled synchronization of wireless networks. The new
methodology can reduce energy consumption in clock synchronization.
QualNet experiments are given to illustrate the analytical results.

\bibliographystyle{unsrt}        
\bibliography{abbr_bibli}           



\balance                                        
\end{document}